\documentstyle[epsfig]{aipproc}
\begin{document}
\title{What Could the Machos Be?\thanks{Based on an invited talk at COSMO-98 in Asilomar, AIP Proc. ``Particle Physics and the Early Universe'', ed. David Caldwell}}
\author{HongSheng Zhao}
\address{Sterrewacht Leiden, Niels Bohrweg 2, 2333CA Leiden, The Netherlands}
\maketitle
\begin{abstract}
If the Universe has a significant baryonic dark component in the form
of compact objects in galaxy halos (machos), then there is a minute
chance (about $10^{-7}$) that one of the Galactic machos passes
sufficiently close to our line of sight to a star out of some $10^7$
monitored stars in the Magellanic Clouds (MCs) that it brightens by
more than $0.3$ magnitude due to gravitational focusing.  After a
brief discussion of the current controversy over the interpretation of
the observed events, i.e., whether the lensing is caused by halo white
dwarfs or machos in general or by stars in various observed or
hypothesized structures of the Clouds and the Galaxy, I propose a few
observations to put ideas of the pro-macho camp and the pro-star camp
to test.  In particular, I propose a radial velocity survey towards
the MCs.
\end{abstract}

{\bf Current Debate.}  Experimental searches for microlenses in the
line of sight to the MCs by MACHO, OGLE, EROS and a number of
follow-up surveys have found more than $16$ candidates to the LMC and
two to the SMC.  Most of them are clustered within $2^o$ of the center
of the LMC (cf. upper left panel of Fig.1).  The observed rate falls
short of explaining the rotation curve of the Galaxy with a smooth
halo of machos (Alcock et al. 1997a).  Whether this is yet another
major puzzle in astronomy which requires explanation by fine tuning is
an open question (Adams \& Laughlin 1996, Chabrier et al. 1996,
Charlot \& Silk 1995, Flynn et al. 1996, Graff \& Freese 1996, Gates
et al. 1998, Gibson \& Mould 1997, Honma \& Kan-Ya 1998).  Another
complication to the otherwise plausible conversion of the event rate
to $\Omega_{macho}$ of the Universe is the inevitable background
events, on top of any macho signal, coming from lensing of two stars
in the LMC disc (Sahu 1994).  Since disc-disc lensing is not very
efficient if the LMC disc is cold and thin (Gould 1994, Wu 1994), more
speculative models have been put forward by the pro-star camp to boost
up the star-star lensing by hypothesizing a variety of special
structures to the MCs.  In particular, a connection is drawn between
the unexpected rate of microlensing and the Milky Way-MCs and SMC-LMC
interactions (Zhao 1998a, b, Weinberg 1998); the latter is yet another
long-standing issue, rooted in the tidal vs. ram-stripping formation
of the Magellenic stream, which is resolved finally by recent
observations (Putman et al. 1998 and references therein).  Several
observations and theoretical arguments suggest that our line of sight
to the LMC passes through a 3-dimensional stellar distribution more
extended in the line of sight than a simple thin disc of the LMC
(Evans et al. 1998, Kunkel et al. 1997, Zaritsky et
al. 1997, 1999, Zhao 1996).  Others argue against additional
structures other than the thin disc of the LMC (Alcock et al. 1997b,
Gallart 1998, Beaulieu \& Sackett 1998, Bennett 1998, Gould 1998,
Johnston 1998, Ibata et al. 1999).  The related issue on the
efficiency of self-lensing of a tidally stretched SMC bar has also
been brought up a few times (Alcock et al. 1997c,
Palanque-Delabrouille et al. 1998, Zhao 1998a), but was highlighted
soon after the discovery of the 98-SMC-1 caustic binary event (Afonso
et al. 1998, Alcock et al. 1999, Albrow et al. 1999, Becker et
al. 1998, Sahu \& Sahu 1998, Udalski et al. 1998).  The observations
are clearly in support of the idea that the SMC has a non-equilibrium
structure extended in the line of sight (Caldwell \& Coulson 1986,
Mathewson, Ford \& Visvanathan 1986, Welch et al. 1987, Westerlund
1997).  While there is still some dispute about the interpretation of
the two binary lens candidates LMC-9 (Bennett et al. 1996, Zhao 1998a)
and 98-SMC-1 (Honma 1999, Kerins \& Evans 1999), the consensus seems
to be that self-lensing in the LMC and SMC is significant if not
dominant.  In general, models of the lens and source distribution to
the LMC fall in two broad classes.

\begin{figure}[b!] 
\vskip -0.5 cm
\epsfysize=5cm 
\epsfxsize=7cm 
\leftline{\epsfbox{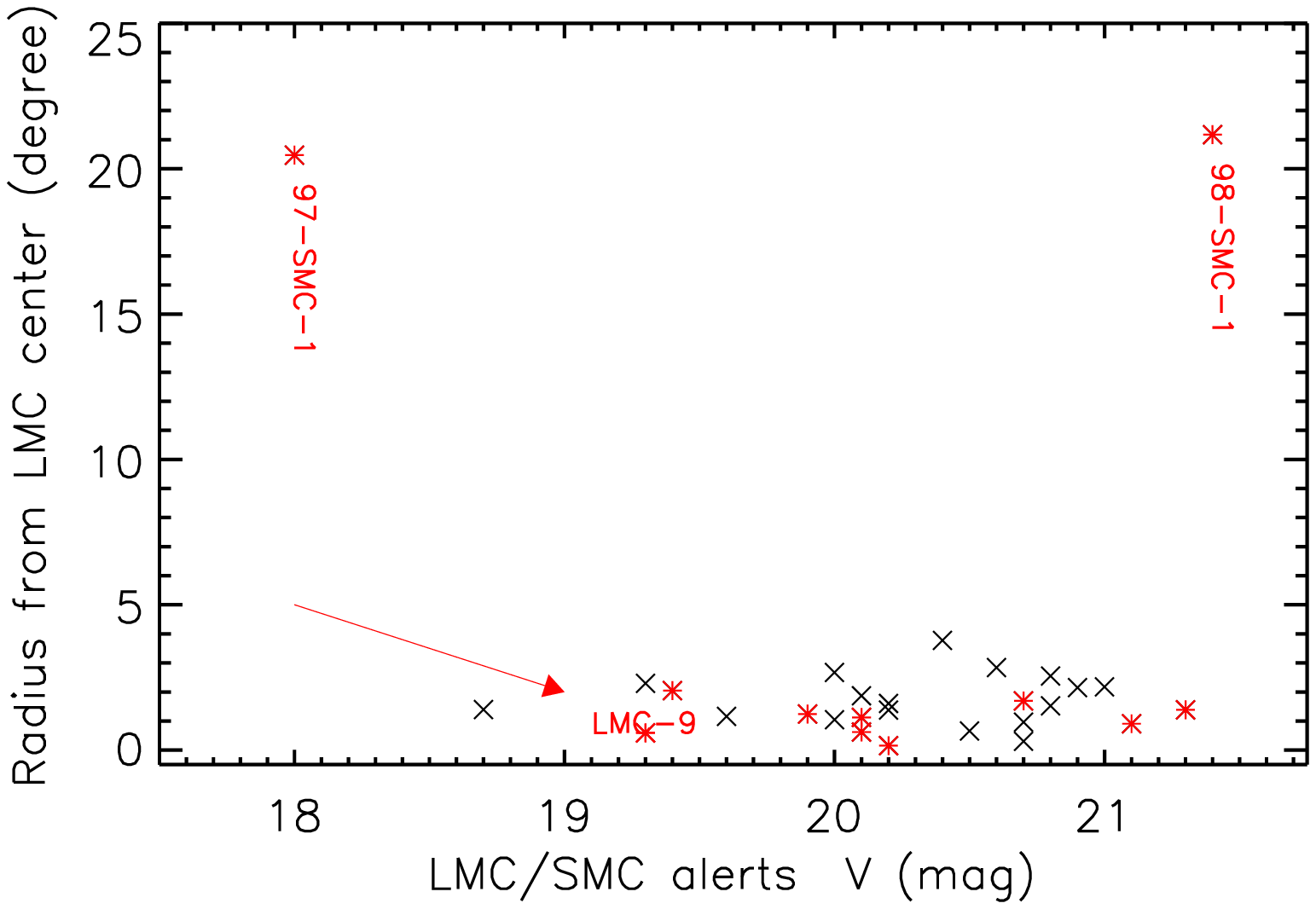}}
\epsfysize=5cm 
\epsfxsize=7cm 
\leftline{\epsfbox{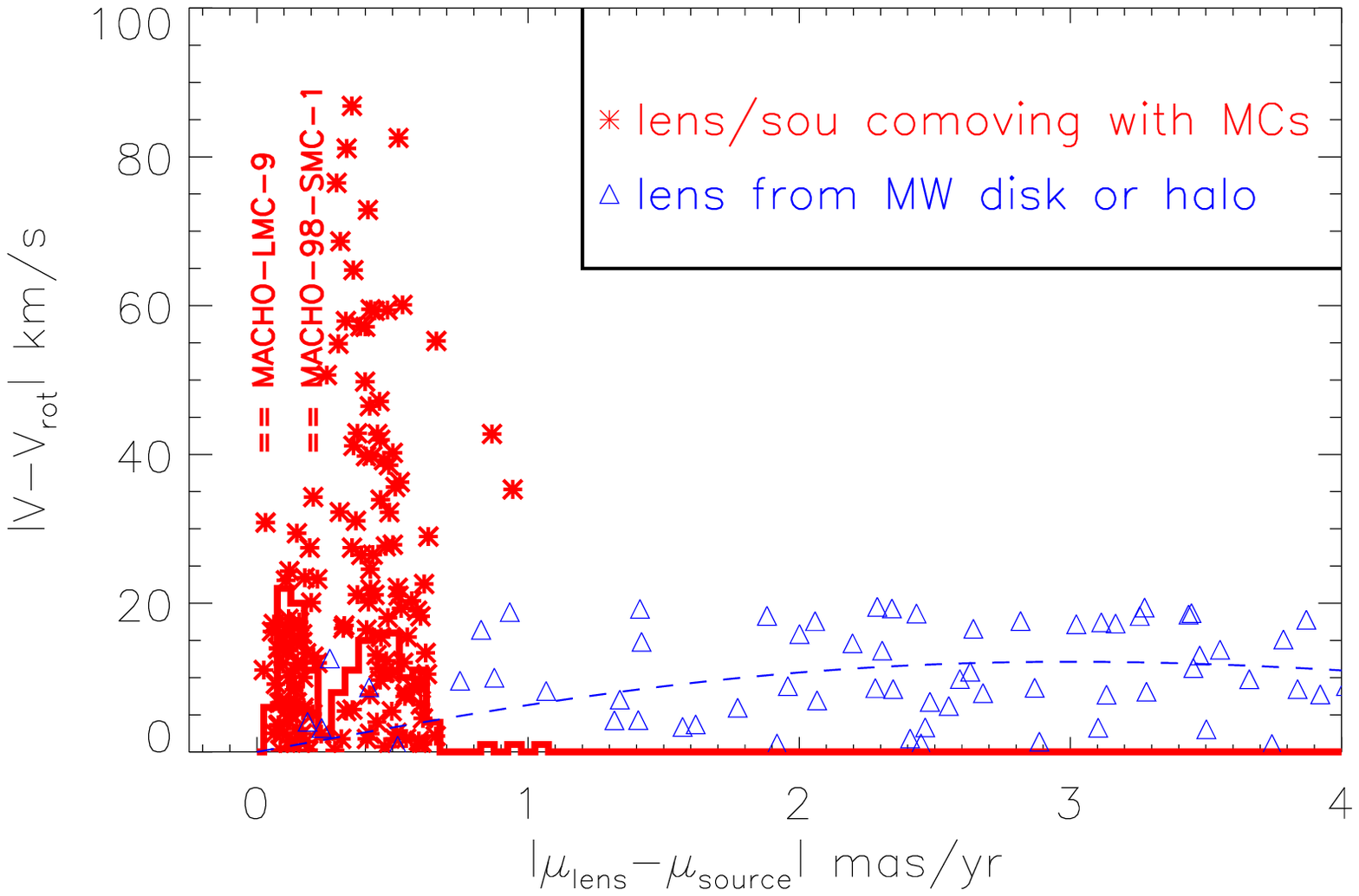}}
\vskip -10cm
\epsfysize=10cm 
\epsfxsize=7cm 
\rightline{\epsfbox{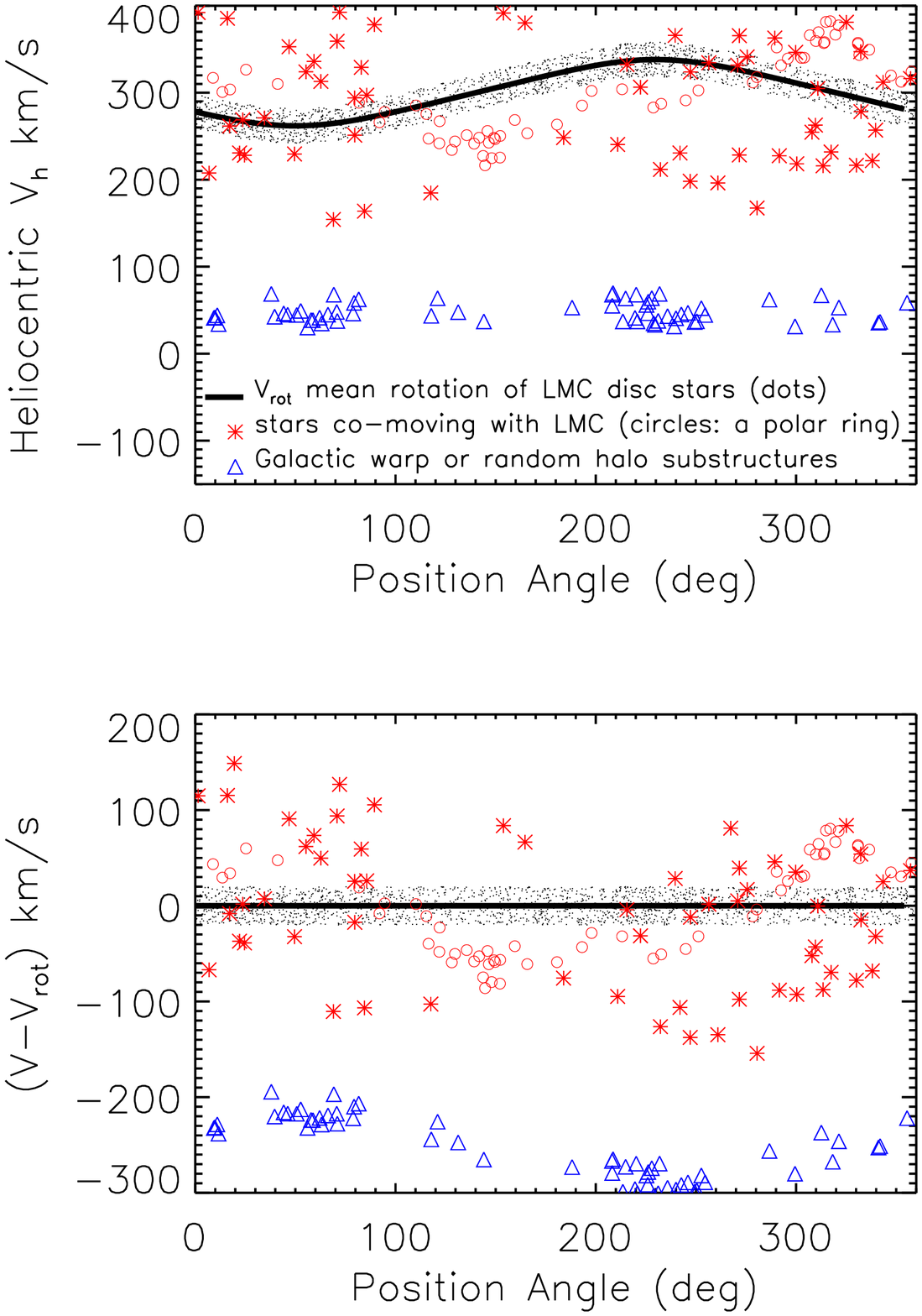}}
\caption{ Upper left: Magnitude and radius distribution of some 30
microlensing alerts (crosses) and published candidates (asterisks)
towards the MCs.  Lower left: Relative source-lens
proper motion vs. the offset in radial velocity of lensed
source from average LMC disc stars in the same field.  Upper right: The radial
velocity of stars in the general direction of the LMC as a function of
the position angle.  Lower right: Distribution of the same objects
after removing the mean rotation of the LMC disc (the sinusoidal
curve). }
\end{figure}

\begin{figure}[b!] 
\vskip -0.5 cm
\epsfysize=9.5cm 
\epsfxsize=7cm
\rightline{\epsfbox{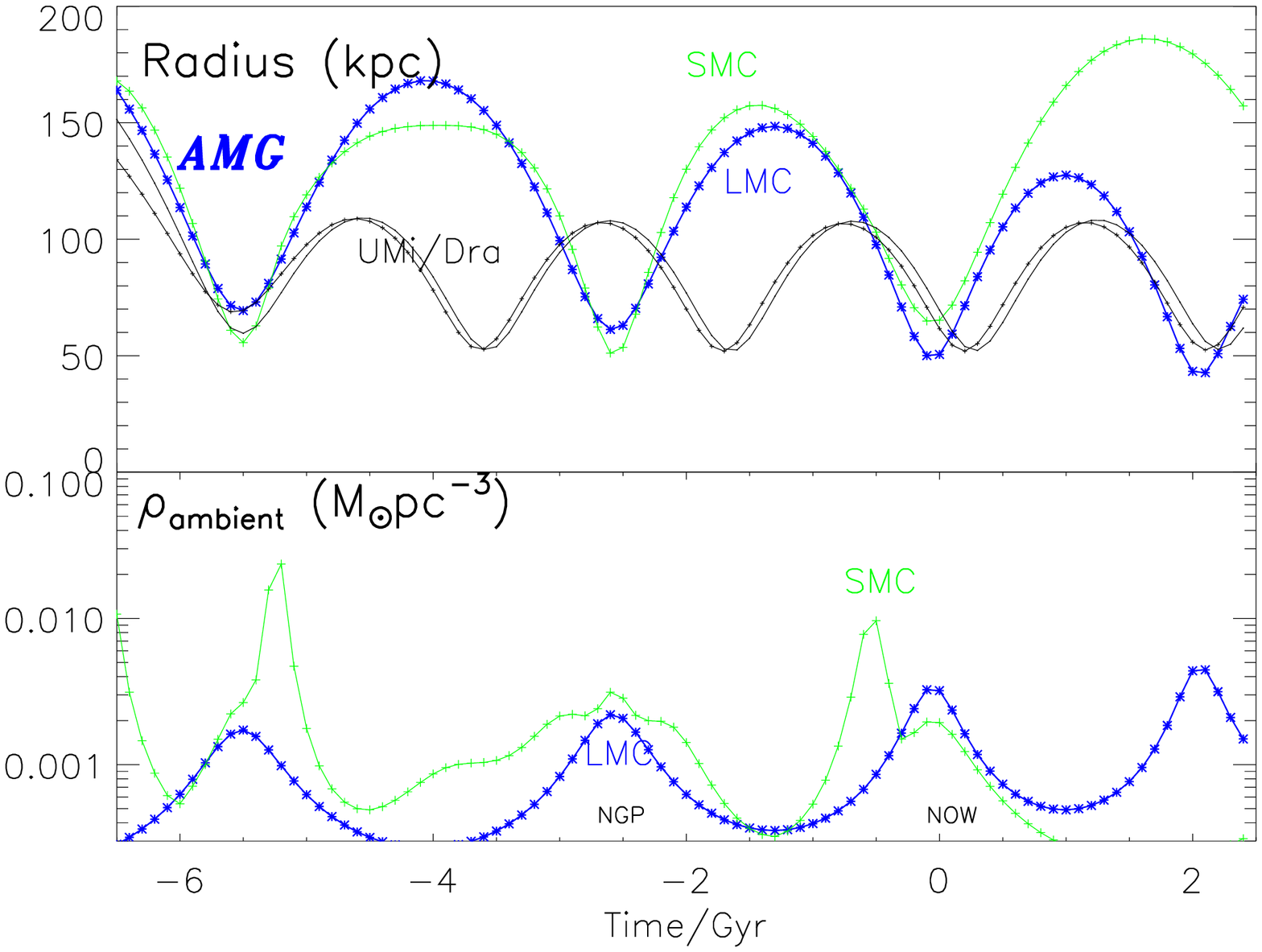}}
\vskip -9.5cm
\epsfysize=5cm 
\epsfxsize=7cm
\leftline{\epsfbox{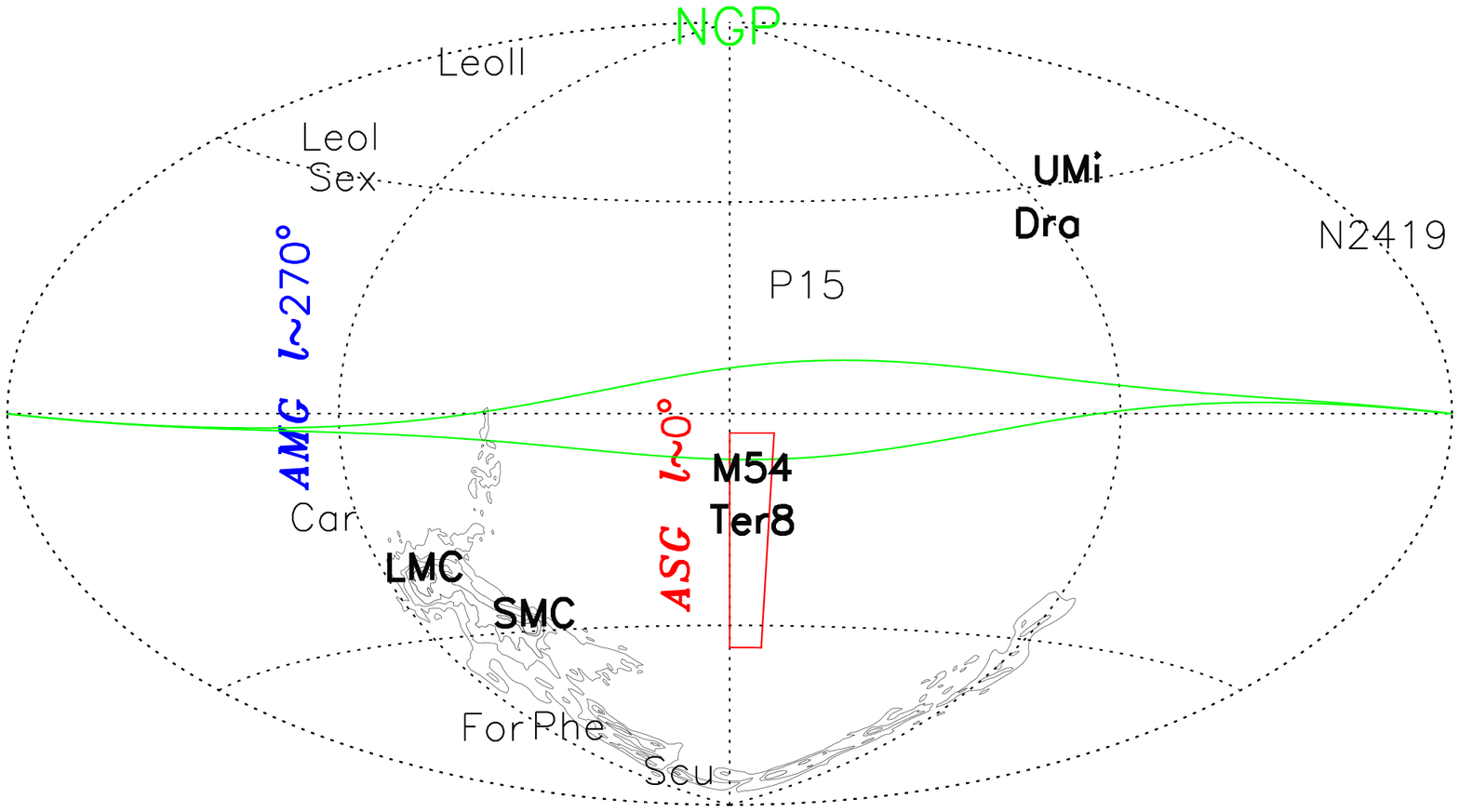}}
\vskip -0.5cm
\epsfysize=5cm 
\epsfxsize=7cm
\leftline{\epsfbox{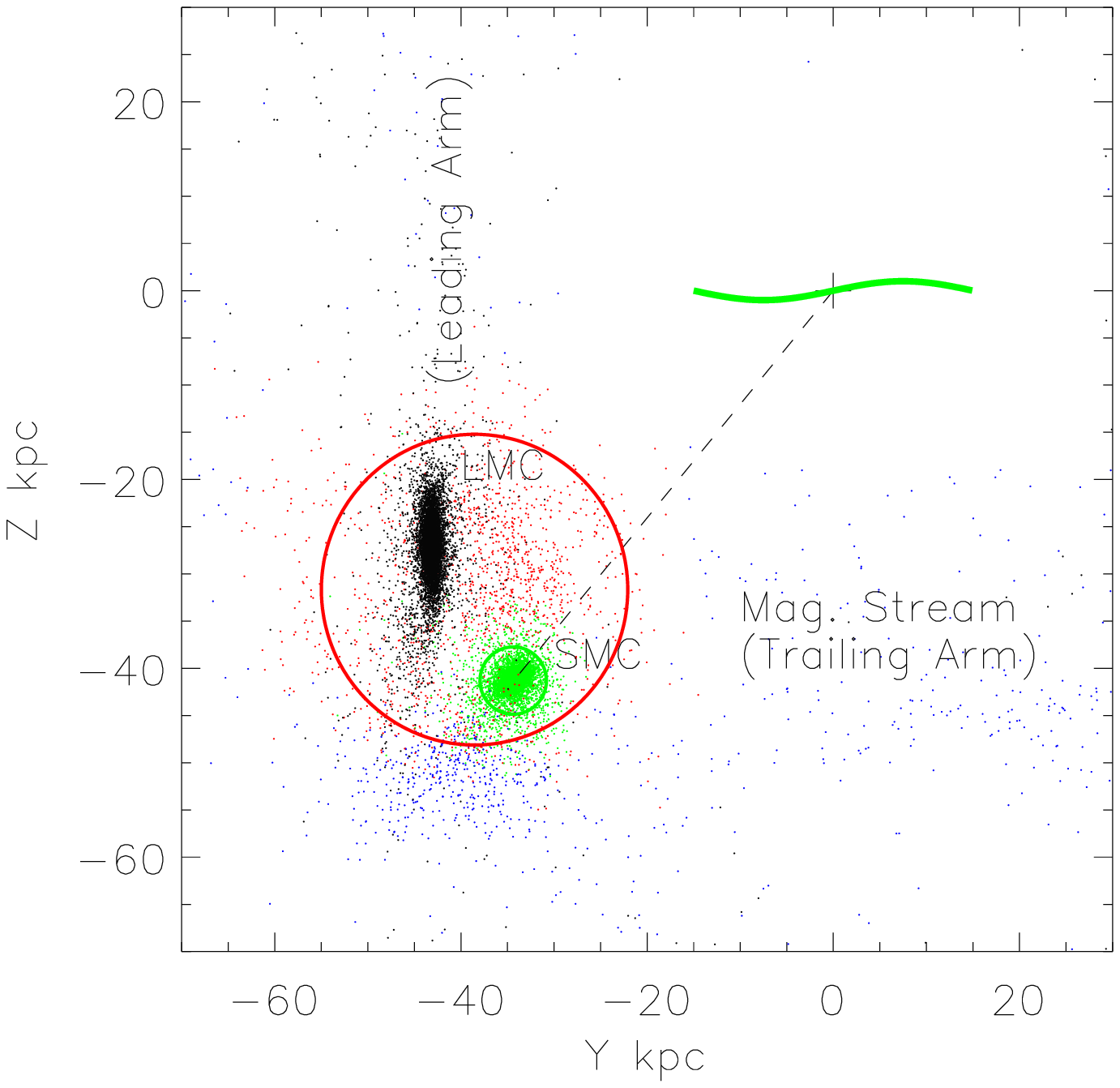}}
\caption{Upper left: Aitoff projections of a few halo dwarf galaxies
and globulars, the Magellanic Stream, the disrupted Ancient
Sagittarius Galaxy (ASG, along $l\sim 0^o$) and a sketch of a warped
disc of the Galaxy. The Ancient Magellanic Galaxy (AMG) is the
suspected common progenitor of objects on the great circle $l\sim
270^o$. Lower left: Simulated tidal structures along the Magellanic
Stream and in the vicinity of LMC/SMC.  The Sun-Galactic center line
is along $Y=Z=0$.  Upper right: Predicted evolution of the distance from
the Galactic center of the LMC, SMC, Ursa Minor and Draco for the past
$6.5$ Gyrs under the assumption that they come from a common
progenitor.  Lower right: Predicted history of tidal shocks (measured
in ambient density) on the MCs, including a very recent one on the
SMC.}
\end{figure}

{\bf Objects with motion decoupled from the Clouds.}
These are dark objects or stars of origin independent of the MCs.  
They generally move with a velocity quite different from the
MCs.  Besides the Galactic dark halo, they include the thick
disc, and maybe a segment of the warp of our Galaxy (Evans et al. 1998)
or even a Sagittarius-like undigested substructure in the halo which
by chance is in the foreground or background of the MCs (Zhao 1996).
Recent observations, including RR Lyrae sample in the MACHO sample,
are not in favor of chance alignment models.  Stars in the warp would
have small heliocentric velocities as they participate in the Galactic
rotation just like the Sun.  The hypothesized substructure should be a
cold feature in the velocity range between $-300{\mbox{\rm \,km\,s}}^{-1}$ 
to $300{\mbox{\rm \,km\,s}}^{-1}$.

{\bf Objects co-moving with the Clouds.}  These are generally stars
which are one way or another generated by dynamical processes in the
formation and evolution of the MCs or their progenitor.  A predicted
orbital and tidal disruption history of the Ancient Magellanic Galaxy
(AMG) is shown (the two right panels of Fig. 2) in a model for the
progenitor galaxy (Zhao 1998b).  The lower left is a plot of the
end-result of an N-body simulation (kindly made available by Lance
Gardiner) of the LMC-SMC-Milky Way interaction, which creates a rich
variety of stellar and gaseous substructures shown here.  Zhao (1996,
1998a,b) suggested lensing of several such substructures, loosely
bound or completely unbound to the MCs.  These include a completely
unbound grand tidal arm of the MCs which extends to Ursa Minor and
Draco, a localized common halo bridging the MCs, a tidally strongly
deformed SMC, a polar ring structure or a tidal halo of the LMC
(cf. Fig. 2).  The LMC disc might also not be perfectly thin and flat.
Similar to the Galactic disc, which has a strong warp at the edge of
the disc due to perhaps tidal interaction with the LMC or the
Sagittarius dwarf galaxy, the gas-rich LMC disc might sport a
thickened disc (Weinberg 1998) due to repeated harassment of the SMC
and the Galaxy.  All these models imply tidal interaction and could
match observations, such as, the Magellanic Stream and its leading
arm, proper motions of Ursa Minor and Draco, distributions of gas and
stellar associations in the Magellanic Bridge, and a possible large
line of sight extent of the SMC and a polar ring structure seen in the
velocity distribution of LMC carbon stars (Kunkel et al. 1997).
Weinberg's model is also motivated by a diffuse structure seen in star
count data of USNO2 which seems to end near the tidal radius of the
LMC.

These extra stars surrounding the Clouds are still highly speculative
with highly uncertain geometries.  It is debatable whether their
structure is better described as a thickened disc, or a warp, a flare,
a polar ring, a tidal halo etc..  But the common feature is that they
circulate around the Galaxy together with the Clouds. In this sense
they can be called co-moving objects, with a proper motion and radial
velocity typically within, say $100{\mbox{\rm \,km\,s}}^{-1}$, of MCs.
For example, the Vertical Red Clump feature identified by Zaritsky et
al. (1997, 1999) in their color-magnitude diagram of the LMC might be
stars a few kpc in front of the LMC disc.  If this interpretation
holds up against countering arguments, the VRC objects would belong to
the co-moving class rather than an independent stream since they move
with a velocity very similar to the LMC.

These models typically predict a lens-source separation $>1$ kpc and
events of diameter crossing time $>30$days for a $0.1M_\odot$ stellar
lens, and require a total mass in tidal substructures above $10\%$ of
that of the LMC, or comparable to that of the SMC to produce
significant lensing.  The challenge is to keep a large amount of
unbound material near the Clouds, unless they are born recently (say
due to a recent SMC-LMC encounter) or they remain loosely bound.  For
example, part of the common halo or bridge in the simulation shown in
Fig. 2 is loosely bound to the MCs.  It contains about $10^9M_\odot$
in a mix of gas and stars with a velocity dispersion of about
$60{\mbox{\rm \,km\,s}}^{-1}$ (Gardiner et al. 1994, 1996).

{\bf Smoking Guns for the two broad classes of models.}
Star-star lensing leaves detectable traces in the kinematical and
spatial distribution of the lensed sources.  This is illustrated
schematically by the right two panels of Fig. 1.  The narrow
sinusoidal band of the PA vs. $V_r$ plot is indicative of the rotation
of the kinematically cold (dispersion less than $20{\mbox{\rm
\,km\,s}}^{-1}$) young thin disc component; an older, thicker disc
would also rotate but with a larger dispersion.  Overplotted are the
predicted radial velocities of extra stars coming from either a
decoupled component (triangles), which is clearly offset from the LMC,
or a substructure comoving with the LMC (asterisks), which would
generally show as a thicker band with either little dependence on the
PA, or dependence different from the rotating disc (e.g., stars on the
polar ring of Kunkel et al.).  The co-moving material can be
identified after subtracting the rotation (the lower right panel);
they are markedly offset from the narrow distribution of the LMC thin
disc stars.  Here I propose a number of ways to resolve the
substructures in the line of sight.

{\bf I.} Self-lensing of stars comoving with the Clouds would induce a
gradient of the event rate per survey star per year across the LMC
disc since it is modulated by the structure of the LMC, whose width
and density vary with the line of sight.  The typical event time scale
varies similarly.  On the other hand if the lenses come from a smooth
macho halo or any extended smooth structure decoupled from the MCs, it
should be nearly the same for all lines of sight to the LMC disc.  A
tricky point here is that the dark halo could have fine structures,
maybe a Sgr-like stream made of dark machos, which could induce a
gradient as well.

{\bf II.} Self-lensing prefers sources at the backside or behind the LMC
disc.  This is because lensing is most efficient if the source is
located a few kpc behind a dense screen of stars, here the LMC disc.
As a result, one should find a slight bias of lensed sources towards
fainter magnitude than average stars in the survey, which are mostly
LMC disc stars.  Together with the spatial bias in (i) self-lensing
should induce a bias in the magnitude vs. radius relation (maybe in
the direction of the arrow in upper left panel of Fig. 1).  No such
bias is expected for macho models.  Again there could be ambiguity if
the stars at the backside of the LMC disc are systematically younger
or older, hence intrinsically brighter or fainter.

{\bf III.} Furthermore, these lensed sources behind the LMC disc are
kinematically different from average stars in the LMC.  This is
perhaps the strongest signal of self-lensing since radial velocities
can be measured so accurately that we can look for a small offset.
There are two complementing effects here.  First a radial velocity
survey should pick up a small fraction of outliers of the rotation
curve of the LMC disc (cf. right panels of Fig. 1), which may belong
to some puffed-up distribution surrounding the LMC disc.  Second the
outliers at the backside of the LMC are more likely picked as lensed
sources, with a velocity set apart from the rotation speed of the LMC
disc {\it in the same field} by typically more than $20{\mbox{\rm
\,km\,s}}^{-1}$ (cf. the lower left panel of Fig. 1).  In contrast, if
all events come from LMC disc stars being lensed by foreground Milky
Way machos or disc stars, then the lensed sources would follow the
motions of average stars in the cold, rotating disc.

{\bf IV.} A continuation and expansion of the MACHO survey hopefully
can yield a small sample of exotic events (e.g., when a source star
passes the caustics of a binary lens) for which we can often tease out
the relative lens-source parallax or proper motion.  Their
distribution would be a direct probe of the dynamics and structure of
the lens population.  This effect can be integrated with the radial
velocity bias to set part various lens populations (cf. lower left
panel of Fig. 1).  The potential has clearly been demonstrated by
LMC-9 and 98-SMC-1, where the observed small relative proper motion
also seems to favor lensing by stars co-moving with the LMC (asterisks
and p.m. histogram) than lensing by foreground populations in the
Milky Way (triangles and the broad Gaussian curve with a
p.m. dispersion of $3$mas/yr, typical for halo machos or Galactic
thick disc stars).  I expect a handful of such events in the next five
years with an extrapolation of current surveys, while the Next
Generation Microlensing Survey (NGMS, Stubbs 1998) holds the promise
of a similar number of events each year.  Again whether a sample of
caustic crossing binary events is a fair sample depends on whether we
expect the fraction of machos in close binaries to be the same as that
of stars.

Detection or non-detection of these subtle effects would be the key to
resolve the controversy of the Galactic dark matter.  The radial
velocity survey is by far the most promising approach since it is not
limited to the rare high amplification (caustic) events, it is not
essential to take spectra of the source star during lensing, and the
effect is big and easy to detect.  While the strength of each of the
``smoking guns'' are still subject to details of our assumptions of,
e.g., the star formation history of the LMC and intrinsic properties
of machos, a complementary studies of lensing-induced systematic bias
in radial velocity, distance, proper motion and spatial distributions
is likely to give a robust conclusion on the nature of Galactic dark
matter.  More detailed simulations of the velocity distributions of
various structures will be reported elsewhere, together with
discussions of optimal strategies of carrying out the proposed
observations to lift current degeneracies in interpreting the
microlensing data.  Finally studies of these effects can all be
integrated in the NGMS.  Tests of these effects could be first done
towards the Galactic bulge where self-lensing of the Galactic bar
surely plays an important role.  New constraints will also come from
microlensing surveys towards M31 (Crotts \& Tomaney 1996, Ansari et
al. 1997).  A more distant but exciting prospect is to complement
ground-based microlensing light curves with astrometric follow-up
observations with the SIM satellite (2005-2010).  Furthermore, GAIA
(under study for launching 2009-2014) could in principle detect
microlensing on the basis of astrometric shift alone without relying
on the alerting system from the ground.  The relative proper motion
and parallax of the lens and the source will be the definitive test of
the lens location and kinematics.

The author thanks Wyn Evans, Ken Freeman, Puragra Guhathakurta,
Rodrigo Ibata, Konrad Kuijken, Joel Primack, Penny Sackett and Chris
Stubbs for helpful discussions, and Dennis Zaritsky for a critical
reading.

{}

\end{document}